# Evidences for a ~60-year North Atlantic Oscillation since 1700 and its meaning for global climate change

Adriano Mazzarella[1]   and   Nicola Scafetta[2,3]

[1] Meteorological Observatory - Department of Earth Science, University of Naples Federico II    S. Marcellino, 10  80138   Naples Italy

[2] Active Cavity Radiometer Irradiance Monitor (ACRIM) Lab. Coronado, CA 92118, USA

[3] Department of Physics, Duke University, Durham, NC 27708, USA

Correspondence to: Adriano Mazzarella (adriano.mazzarella.unina.it)

## Abstract

The North Atlantic Oscillation (NAO) obtained using instrumental and documentary proxy predictors from Eurasia is found to be characterized by a ~60-year dominant oscillation since 1650. This pattern emerges clearly once the NAO record is time integrated to stress its comparison with the temperature record. The integrated NAO (INAO) is found to well correlate with the Length of the Day (LOD) (since 1650) and the global surface sea temperature (SST) record (since 1850). These findings suggest that INAO is an excellent proxy for global climate change, and that a ~60-year cycle exists in the global climate since at least 1700. Finally, the INAO ~60-year oscillation well correlates with the ~60-year oscillations found in the historical European aurora record since 1700, which suggests that this ~60-year dominant climatic cycle has a solar-astronomical origin.

# 1 Introduction

Understanding climate change and its causes is a current major scientific and social issue. General circulation models (GCMs), which utilize the known fundamental principles of physics, are commonly used to interpret climate changes and global warming since 1850 (IPCC, 2007). These studies indicated that anthropogenic forcing is the predominant contributor to climatic changes observed during the last century.

However, the results are somewhat controversial because the GCM simulations fail to reproduce the oscillations found in the climate system at multiple scales since 1850 (Scafetta, 2010). In particular, the climate system appears to be characterized by a large ~60-year oscillation that is not reproduced by GCMs. This 60-year oscillation was in its warm phase from the 1970s to about 2000 and could explain a significant part of the observed global warming during the same period. Scafetta (2010) argued that because climate oscillations are coherent with astronomical cycles at multiple time scales, the current GCMs do not include important astronomical forcings of the climate. Consequently, natural causes of climate changes may be still severely underestimated.

However, 160 years of global surface temperature data (since 1850) may be not sufficient to demonstrate that the climate is characterized by a major ~60-year oscillation. Longer global climate records should confirm this claim, but instrumental global surface records before 1850 do not exist.

Indeed, several local climatic proxy records as long as thousand years present a dominant 50-70 year large oscillation. Some of these examples include ice core sample, pine tree samples, sardine and anchovy sediment core samples, global surface temperature records, atmospheric circulation index, length of the day index, fish catching productivity records, etc. (Klyashtorin et al., 2009; Mazzarella, 2007; 2008; 2009). A dominant 60-year periodicity is found in instrumental temperature reconstructions since 1500 covering the European Mediterranean Basin (Spain, France and Italy) (Camuffo et al., 2010). A ~60-year periodicity is found in secular monsoon rainfall records from India, from Arabian Sea sediments and from over east China (Agnihotri and Dutta, 2003). A 60-year oscillation is found in the G. Bulloides abundance variation record found in the Cariaco Basin sediments in the Caribbean sea since 1650 (Black et al., 1999; Scafetta, 2010). This record is an indicator of the trade wind strength in the tropical Atlantic Ocean and of the North Atlantic Ocean atmosphere variability. Other proxy records for the Atlantic Multidecadal Oscillation (AMO) have been found to present a ~60-year oscillation during the entire Holocene for thousand years (Knudsen et al., 2011). A significant 60-year oscillation is found in a reconstruction of global sea level since 1700 calculated from tide gauge records (Jevrejeva et al., 2008). Moreover, Patterson el al. (2004) found 60-62 year cycles in sediments and cosmogenic nuclide records in the NE Pacific. A ~60-year cyclicity has been found in several other solar related records (Ogurtsov et al., 2002; Komitov, 2009; Scafetta, 2010). The above results clearly suggest the existence in the climate of an astronomically induced ~60-year oscillation.

However, that several "local" climatic proxy records may present ~60-year oscillations may be still unsatisfactory. It is necessary to determine whether a ~60-year oscillation characterizes the climate reconstruction of at least extended regions of the Earth that can be used as more reliable proxies for the global climate.

Several multisecular paleoclimatic multi-proxy reconstructions of the global climate have been proposed (North et al., 2006), but they appear quite different from each other and are characterized by very large errors. Nevertheless a significant 50-80 year oscillation is normally found in these paleoclimatic proxy temperature reconstructions (Ogurtsov et al., 2002). However, errors in the data, physical complexity of the proxy records and volcano eruptions can easily disrupt and, consequently, hide a more regular pattern. On the contrary, the clear quasi 60-year oscillation already found in the global sea level record since 1700 (Jevrejeva et al., 2008) does suggest that this pattern is global.

Herein we find further confirmation that the global climate is characterized by a ~60-year oscillation by focusing our analysis on the monthly North Atlantic Oscillation (NAO) multisecular reconstruction proposed by Luterbacher et al. (1999, 2002) since 1659. The North Atlantic Oscillation (NAO) is the main synoptic mode of atmospheric circulation and climate variability in the North Atlantic/European sector and has a substantial influence on marine and terrestrial ecosystems and regional socio-economic activity. This NAO reconstruction (see Figure 1) was obtained using up to 110 Eurasian predictors, which include up to 60 instrumental station pressure predictors and many other precipitation, temperature, tree ring and ice core records.

Moreover, we use the historical Length of the Day (LOD) (Stephenson and Morrison, 1995) and the global instrumental sea surface temperature (SST) record (Brohan et al., 2006) to argue that the time-integrated record of the North Atlantic Oscillation (INAO) is a reliable global climate proxy. Finally, we compare the INAO oscillations with those observed in the European historical record of middle latitude aurorae (Krivsky and Pejml, 1988) to claim that a ~60-year oscillation exists in the global climate and likely has an astronomical origin as previously proposed (Scafetta, 2010).

## 2  Collection of data

We analyzed the historical series of:
a) North Atlantic Oscillation (NAO) (hPa) defined as the normalized sea level pressure between the Azores high and Icelandic low (Rogers, 1984; Hurrell, 1995; Jones et al., 1997) (interval, 1861-2009). Due to the brevity of the available instrumental records, we used the series of yearly NAO values as reconstructed by Lutherbacher et al. (1999; 2002) who utilized instrumental and documentary proxy data covering the entire Eurasia to extend NAO back to 1659. The yearly data of reconstructed NAO are taken from the web site:
http://www.esrl.noaa.gov/psd/gcos_wgsp/Timeseries/RNAO/
and reported in figure 1. For convenience, we integrate NAO yearly values into an integrated NAO (INAO) annual record according to a sequential summation of NAO, i.e., INAO(t) =INAO(t-1) + NAO (t) for each year "t". Figure 2a depicts INAO.

b) The Earth's rotation speed, as normally measured by means of LOD (ms). This record represents the difference between the astronomical LOD and the standard length (interval: 1657–2009) (Stephenson and Morrison, 1995). The LOD yearly data are taken from the web site:
http://hpiers.obspm.fr/eop-pc/earthor/ut1lod/lod-1623.html
and reported in figure 2b.

c) Global sea surface temperature SST anomaly (°C) provided by the Climatic Research Unit, University of East Anglia (interval: 1850–2010) (Brohan et al., 2006; Rayner et al., 2003). We use the SST record as a global index of climate change that may be naturally linked to the NAO. The yearly SST data are available from the web site:   http://www.cru.uea.ac.uk/cru/data/temperature/hadsst2gl.txt
and reported in figure 2c.

d) The annual frequency of auroras observed in Europe since 1700. This record contains the historical aurora observations reported in Europe from 1000 to 1900 AD (Krivsky and Pejml, 1988). Before 1700 the record is largely incomplete and those data are not considered here. The record is merged with the catalog of the aurora observations in the Faroes Islands from 1872 to 1966. The Faroes' record is preferred here because of its length and because it shows physical properties in continuity with the Krivsky and Pejml's record (Silverman, 1992). The two combined catalogs cover 267 years from 1700 to 1966. The yearly aurorae frequency data are available from the web site:
ftp://ftp.ngdc.noaa.gov/STP/SOLAR_DATA/AURORAE/
and are depicted in figure 2d.

## 3  Analysis and results

The Northern hemisphere zonal circulation is well represented by the NAO (hPa) index, that is a gradient of normalized atmospheric pressure ($Nm^{-2}$), directly proportional to an acceleration ($s^{-2}$). To directly compare the forecasting reliability of NAO and LOD as proxy for climatic changes, we integrate NAO yearly values into INAO annual record according to a sequential summation of NAO.

We investigated the cross-correlations between INAO, LOD, SST and aurorae yearly series on different time scales. We linearly normalize to a mean equal to zero and to a standard deviation equal to one each record within a given time scale and utilize the raw normalized records as well as 5-yr, 11-yr and 23-yr running means.

The running mean method is equivalent to a low pass filtering with no change in amplitude and phase (Bath, 1974) that helps to remove the oscillations shorter than the time scale of the running mean. Such a methodology provides an accurate low-frequency spectral analysis of the investigated series.

Figure 3 reports the time plots of raw yearly values of INAO and LOD and smoothed according to 5-yr, 11-yr and 23-yr running means together with the relative correlation coefficient. INAO is well inversely correlated to LOD ($0.73<R<0.82$), indicating that an increase in zonal wind speed is responsible for an ever more effective decrease in LOD as the time scale increases.

Figure 4 reports the time plots of raw yearly values of LOD and SST and smoothed according to 5-yr, 11-yr and 23-yr running means together with the relative correlation coefficient. The figure demonstrates that LOD is inversely correlated to SST (correlation coefficient: $0.58<R<0.92$). The best correlation is obtained by shifting LOD ahead by 4 years ($0.66<R<0.96$).

Figure 5 reports the time plots of raw yearly values of INAO and SST and smoothed according to 5-yr, 11-yr and 23-yr running means together with the relative correlation coefficient. The figure demonstrates that INAO is directly correlated to SST (correlation coefficient: $0.61<R<0.89$). The best correlation is obtained by shifting INAO ahead by 4 years ($0.74<R<0.96$).

Figure 6 reports the time plots of raw yearly values of INAO and yearly aurora frequency and smoothed according to 5-yr, 11-yr and 23-yr running means. The figure suggests that INAO is inversely correlated to the auroras although the correlation appears poor at first sight. In fact, the dominant pattern in the aurora record is the very large decadal oscillations, which are related to the 11-year solar cycle. Very large decadal oscillations are typical of upper-atmosphere records (Lean, 2005). However, a decadal cycle, although present, is not as large in the records referring to the troposphere or to the surface (Gleisner and Thejll, 2003; Scafetta, 2009). The good correlation between the two records is better manifest in the multi-decadal smooth records of figure 6d.

Figure 7 reproduces the records depicted in figure 6d. Each record is fit with a sinusoidal curve. For the aurorae record the periods of the oscillations are $T = 60 \pm 8$ year and can be well fit with a sinusoidal curve with a period of 61 years. For INAO the period of the oscillations are $T = 63 \pm 10$ year and can be well fit with a sinusoidal curve with a period of 63 years. The figure demonstrates that these two records present a coherent ~60-year oscillation at least since 1700: the correlation coefficient between the two sinusoidal curves is $R = 0.94$.

The high correlation coefficients computed for each pair of the investigated variables indicate that the variables of each pair are statistically inter-related at a level of confidence no less than 95% (Bath, 1974).

# 4 Discussion and conclusions

Determining whether the climate is oscillating at quasi-regular periods is fundamental both for climate forecasting and for better understanding the causes of climate changes. For example, if climate were regulated by large multidecadal cycles correlated to known astronomical cycles, the climate would be strongly influenced by astronomical natural phenomena and could be more easily forecasted. The research should then focus on investigating the specific solar-astronomical causes of climate change.

Scafetta (2010) showed that the global surface temperature records since 1850 present oscillations (including a large ~60-year oscillation) that well cross-correlate with measurable astronomical oscillations, which are linked to planetary motion. As explained in the Introduction, several local proxy climatic records (some of them are as long as several century and millennia) present a dominant 50-70 year oscillation (see for example Klyashtorin et al., 2009 and Knudsen et al., 2011). However, to be credible and avoid possible criticisms, this ~60-year oscillation should also be found in climatic composites that are indicative of extended regions. Global climate paleoclimatic reconstructions are still too uncertain (North et al., 2006) and their uncertainty may mask a clearer ~60-year oscillation, if this oscillation does exist in the climate. Thus, there is the need to find an acceptable compromise.

Herein we have chosen to investigate the North Atlantic Oscillation (NAO) reconstruction by Luterbacher et al. (1999, 2002) which is available since 1659. This record is made of many of the best long-term instrumental data series (pressure, temperature, precipitation) and other high resolution documentary evidences available since the 17$^{th}$ century. This record covers the whole Eurasia, as figure 1 shows, and it can be considered an excellent proxy for the climate variation in the Eurasia region and the north Atlantic region.

We have found that the integrated-NAO (INAO) well cross-correlates with two global indexes: the Length of the Day (LOD) and the global see surface temperature (SST). The ~60-year oscillation found in the INAO would also well correlate to the ~60-year oscillation found in the global sea level record since 1700 calculated from tide gauge records (see figure 3b in Jevrejeva et al., 2008). This finding suggests that INAO can be considered a very good proxy for the global climate and that a ~60-year oscillation is a global feature of the climate system.

LOD is a good proxy for climatic changes under the assumption that it is the integral of the different circulations that occur within the ocean-atmosphere system both along latitude (zonal circulation) and longitude (meridional circulation) (Mazzarella 2007; 2008; 2009). The 1785-1875 interval appears to have experienced strong zonal circulation while the 1700-1775 interval is characterized by weak circulation (fig. 3). It is herein also confirmed that LOD is inversely related to SST (fig. 4). INAO is found to be directly related to SST. Moreover, a peak cross-correlation is found when INAO is shifted ahead by 4 years. If this finding is not due to the limitation of the data, it indicates that an increase in zonal wind speed is responsible for an increase in SST and may be used for forecast temperature.

Indeed, there is a quasi-equilibrium between zonal and meridional circulation: strong zonal circulations cause the contraction of the circumpolar vortex and an increase in air temperature while weak zonal circulations or, equivalently, strong meridional circulations with meandering or cellular patterns cause an expansion of the circumpolar vortex and a decrease in air temperature. Zonal epochs correspond to periods of global warming and meridional ones to periods of global cooling (Lamb, 1972; Lambeck, 1980; Mazzarella, 2007; 2008; 2009).

INAO values are found to be inversely related to those of LOD such that periods of increasing zonal wind speed are accompanied by periods of the Earth's increasing rotational rate while periods of decreasing zonal wind speed are accompanied by periods of the Earth's decreasing rotational rate (Lambeck, 1980). On the other hand, the Medieval Warm Period known also as Medieval Climate Optimum was caused by a persistent and large positive NAO (Trouet et al., 2009) that corresponds to high values of INAO.

Moreover, almost five large ~60-year cycles in INAO have been identified since 1700. The INAO oscillations are well correlated to a ~60-year oscillation found in the historical European aurora record since 1700. Aurorae are clearly solar and astronomical induced phenomena (Komitov, 2009) and are an index of the electrification of the ionosphere. The atmospheric global electric circuit has been proposed to regulate the microphysics of cloud formation (Tinsley, 2008; Enghoff et al., 2011), which can then regulate the global climate through albedo oscillations. Because the albedo oscillations would share the same frequencies of the mid-latitude aurora records, the albedo too would present ~60-year oscillations. Thus, the incoming solar radiation warming the troposphere and reaching the surface would present ~60-year cycles that would force an ocean-atmosphere circulation pattern synchronized to a ~60-year cycle perhaps through the modulation of the Intertropical Convergence Zone (ITCZ) and Azores–Bermuda high-pressure system. In fact, the ~60-year oscillation appears quite strong and regular in multi-millennial proxy records such as those used to reconstruct the Atlantic Multidecadal Oscillation (Knudsen et al., 2011).

The primary physical cause of a natural ~60-year cycle is not yet clearly understood. However, Scafetta (2010) has shown that the solar system oscillates because of the movement of the planets. The Jupiter-Saturn system produces clear ~60-year oscillations in the barycentric movement of the Sun due to its ~60-year tri-synodic period and induces equivalent oscillations in the earth's orbit too. In particular the speed of the Sun relative to the center of mass of the solar system presents clear ~20 and ~60 year oscillations in perfect phase with the ~20 and ~60-year oscillations observed in the climate system (Scafetta, 2010). Moreover, the two major tidal cycles at 9.3 years (Jupiter and Saturn alignment) and 11.86 years (Jupiter period) beat at about 61 years. Thus, a possibility is that the solar activity and/or the heliosphere may synchronize to a natural solar system ~60-year cycle and, through a nuclear amplification mechanism, may make this signal sufficiently macroscopic to influence the Earth's climate too probably through a modulation of cosmic ray which would modulate the cloud cover (Enghoff et al., 2011). Indeed, a mechanism in which the planets of the solar system can affect the sun has been recently proposed (Wolff and Patrone, 2010).

In conclusion, the findings of this work indicate that the global climate likely presents a ~60 year oscillation since at least 1700. This natural oscillation was in its warm phase during the period 1970-2000 and has likely largely contributed to the global warming during this period. Scafetta (2010) evaluated that about 60% of the warming observed since 1970s could be associated to a 60-year oscillation. Finally, this quasi 60-year oscillation likely has a solar-astronomical origin, in agreement with the hypothesis advanced by Scafetta (2010).

**Appendix: Repeating the calculations with HadSST3**

Very recently, after the submission of the present work, the Met Office Hadley Centre has published in its web-site a major adjustment of its global sea surface temperature data set, which is called HadSST3 (Kennedy et al., 2011a, 2011b). The current HadSST3 data set runs from 1850 to 2006 (http://www.metoffice.gov.uk/hadobs/hadsst3/).

The major difference between HadSST2 and HadSST3 is in a major adjustment of the data during period 1940 to 1970. During this period in HadSST3 it is better observed a gradual downward trend (see Figure 8a), which is in an even better agreement with the 60-year cycle cooling phase claimed in this paper to exists in the climate system.

In Figures 8bcd we show the standardized 5-11-23-year moving average curves of the HadSST3 against the correspondent LOD and INAO curves as done in Figures 4bcd and 5bcd, respectively. The figures show that also with this adjusted SST record the extremely good correlation (b. R=0.76-0.75; c. R=0.83-0.84; d. R=0.89-0.92) of a quasi 60-year modulation between the global sea surface temperature with the LOD and INAO exists. The cross-correlation is even slightly better, because of a better agreement during the period 1940-1970. Thus, the major conclusion of this paper would not be altered but, on the contrary, it would be further confirmed by this new global sea surface temperature record.

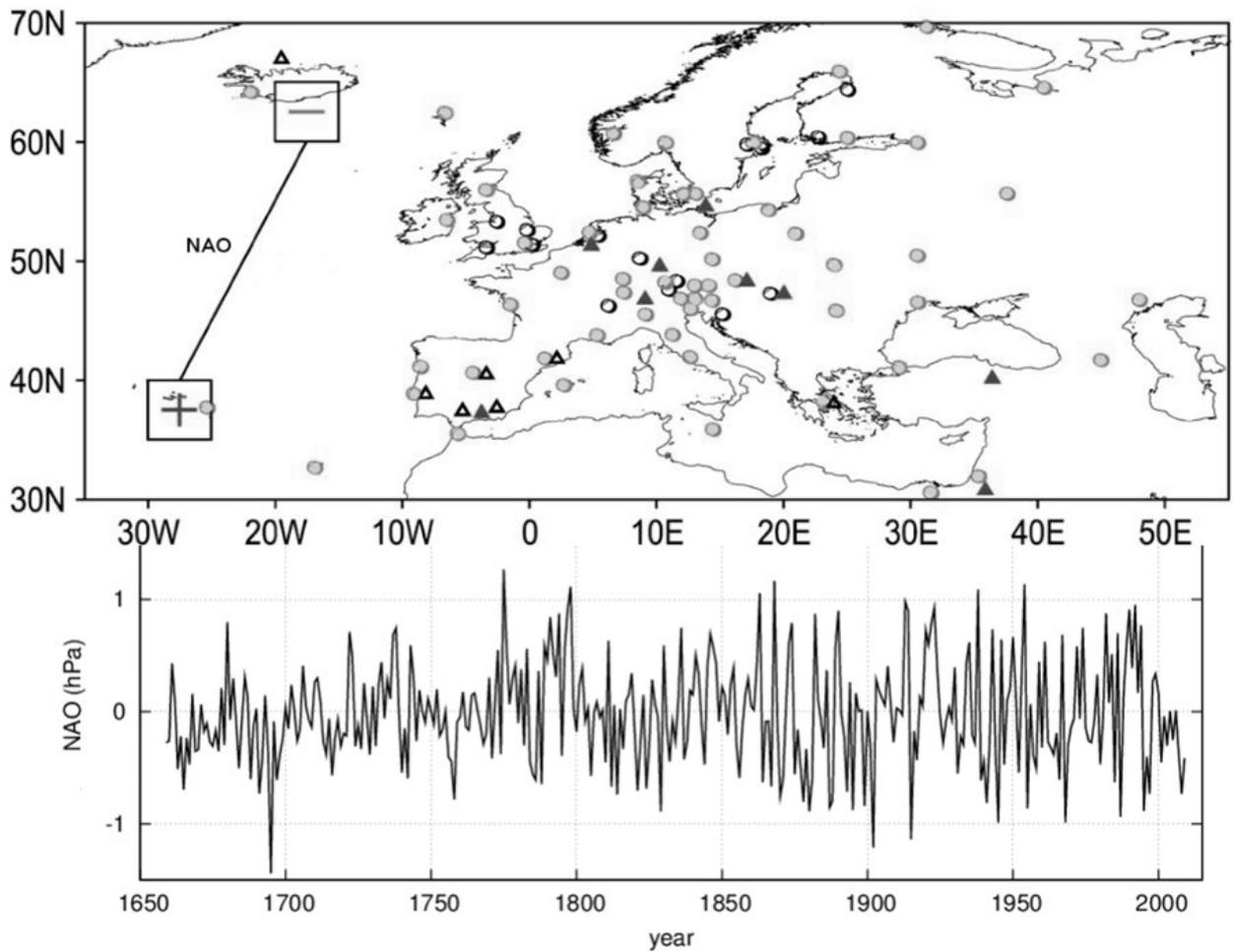

Figure 1. Time plot of yearly values of reconstructed North Atlantic Oscillation (NAO). The upper figure defines the NAO and indicates the locations of the predictors used for reconstructing NAO. For detailed information, see Luterbacher *et al.* (1999, 2002).

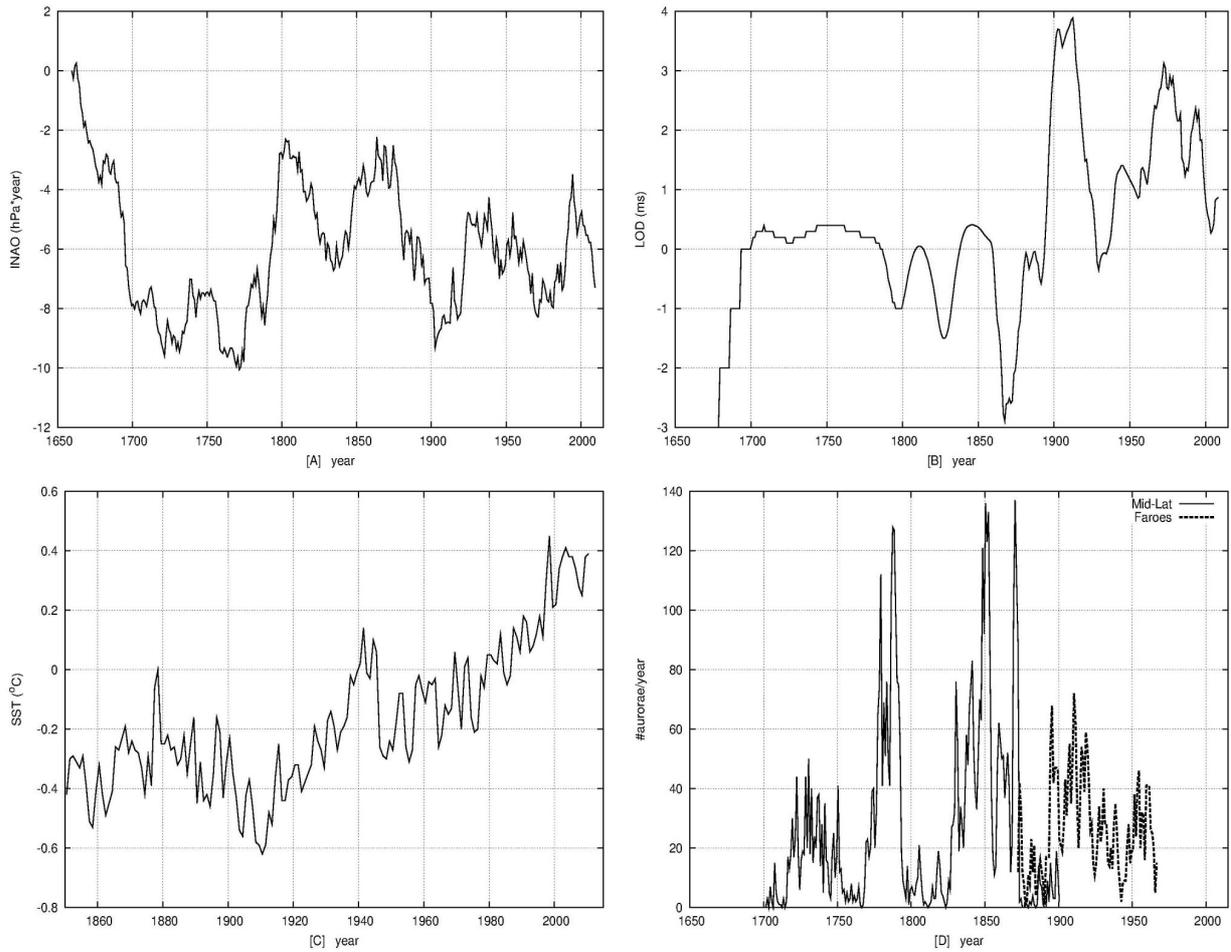

Figure 2. Time plot of yearly values of: (a) integrated NAO oscillation INAO. ; (b) length of day LOD; (c) sea surface temperature SST; (d) yearly frequency of European aurorae. See text for references.

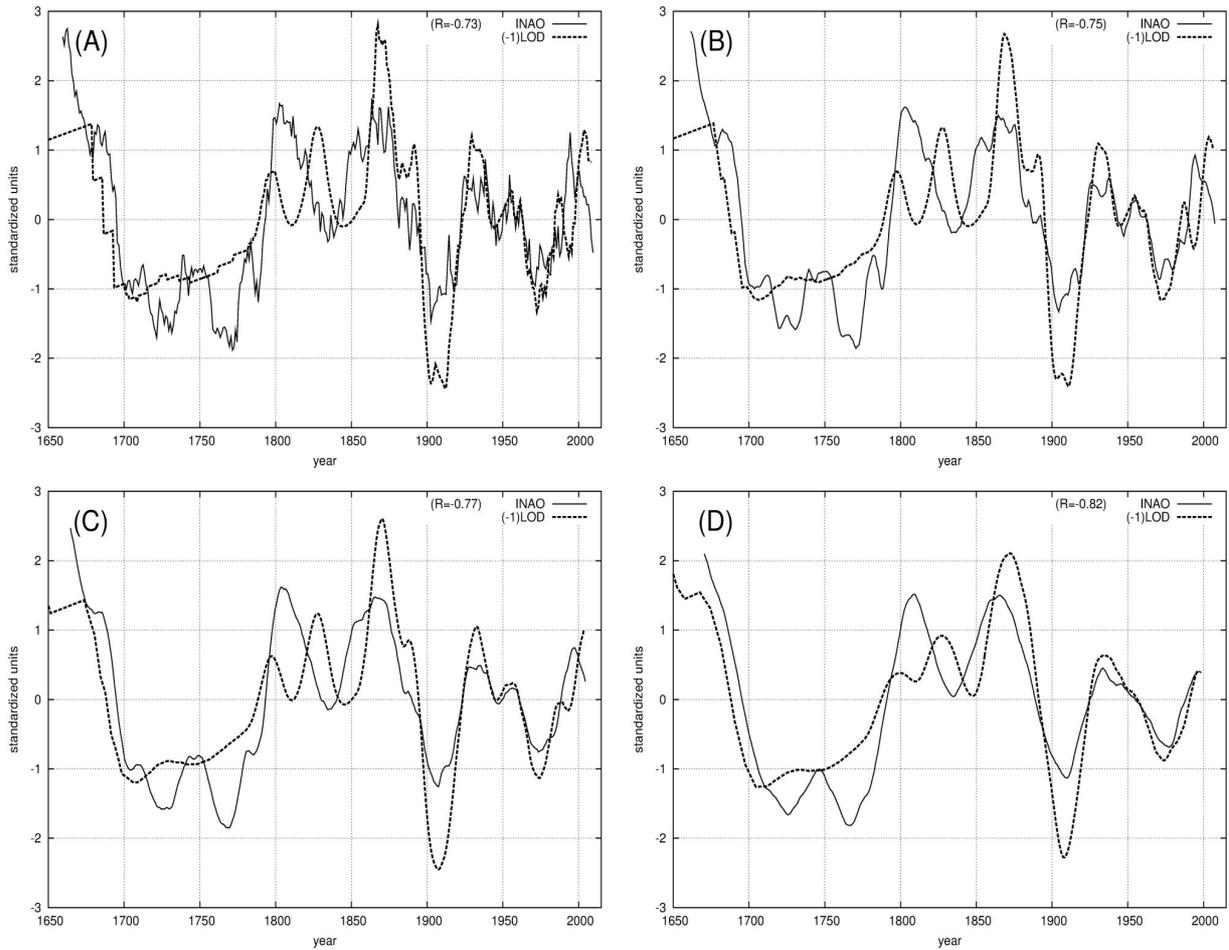

Figure 3. Time plot of detrended yearly values of INAO and LOD: a) raw values, b) smoothed according to a 5-yr running mean, (c) smoothed according to a 11-yr running mean, d) smoothed according to a 23-yr running mean. The symbol (-1) before LOD indicates that the record has been up-down flipped for a better visual comparison.

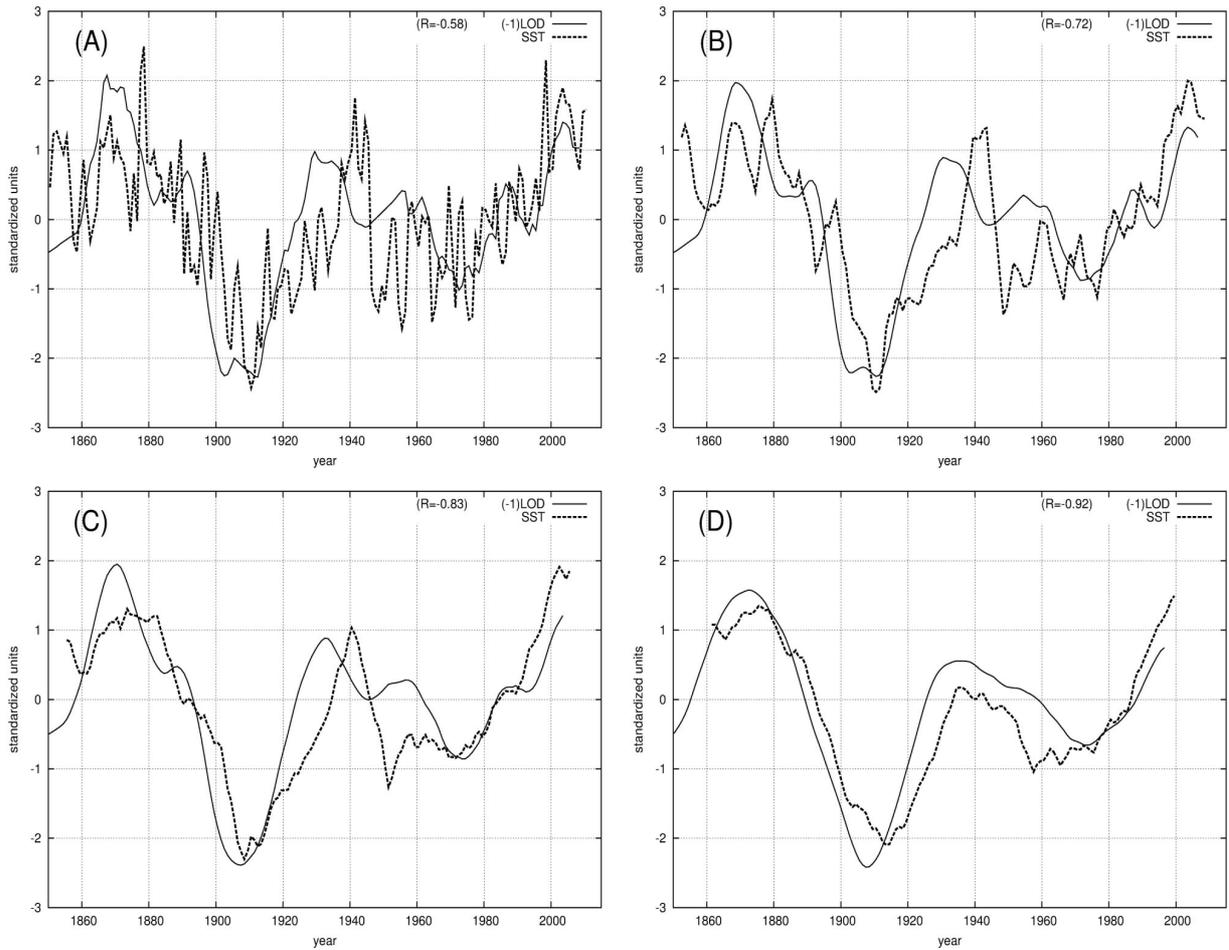

Figure 4 . Time plot of detrended yearly values of LOD and SST: (a) raw values, b) smoothed according to a 5-yr running mean, (c) smoothed according to a 11-yr running mean, (d) smoothed according to a 23-yr running mean. The symbol (-1) before LOD indicates that the record has been up-down flipped for a better visual comparison.

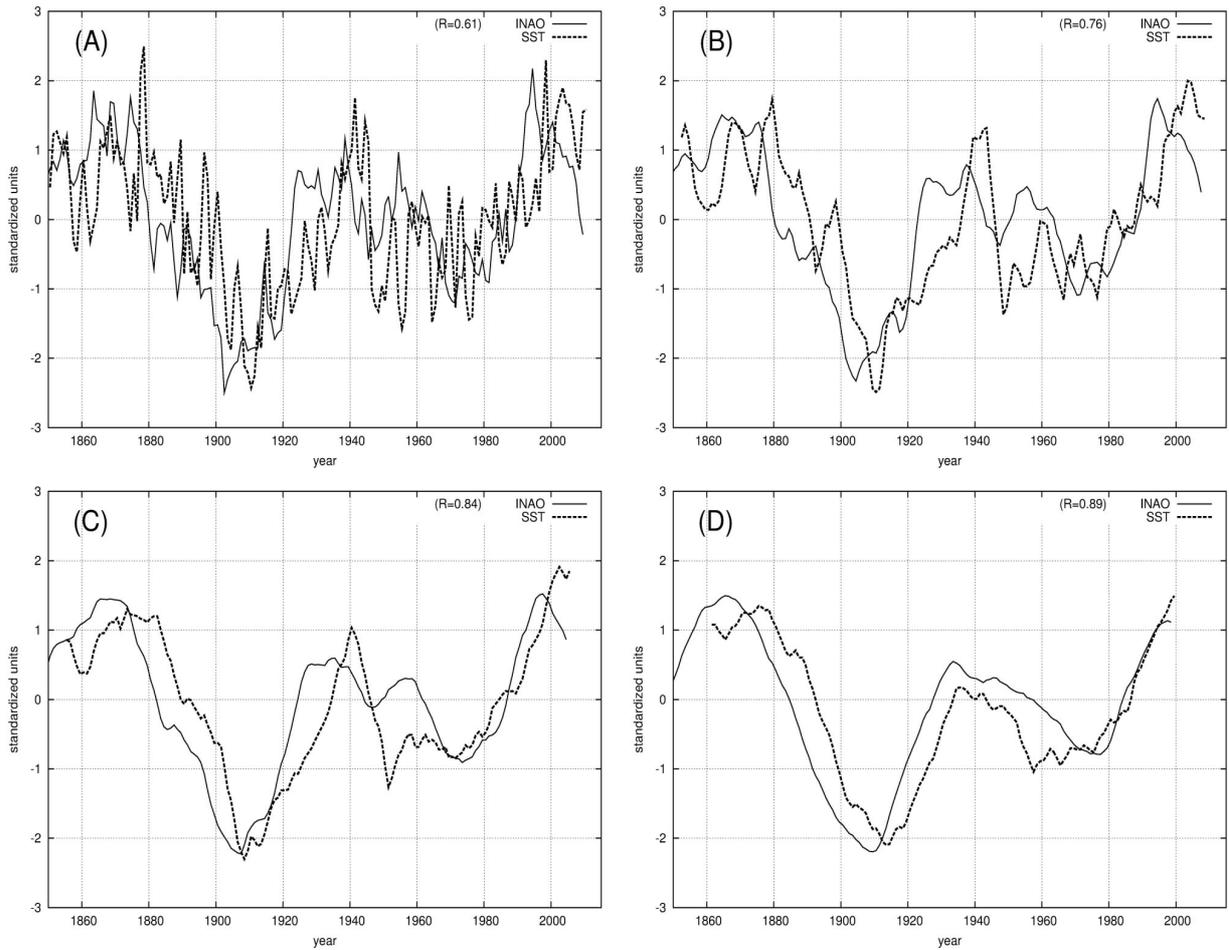

Figure 5. Time plot of yearly detrended values of INAO and SST: (a) raw values, (b) smoothed according to a 5-yr running mean, (c) smoothed according to a 11-yr running mean, (d) smoothed according to a 23-yr running mean.

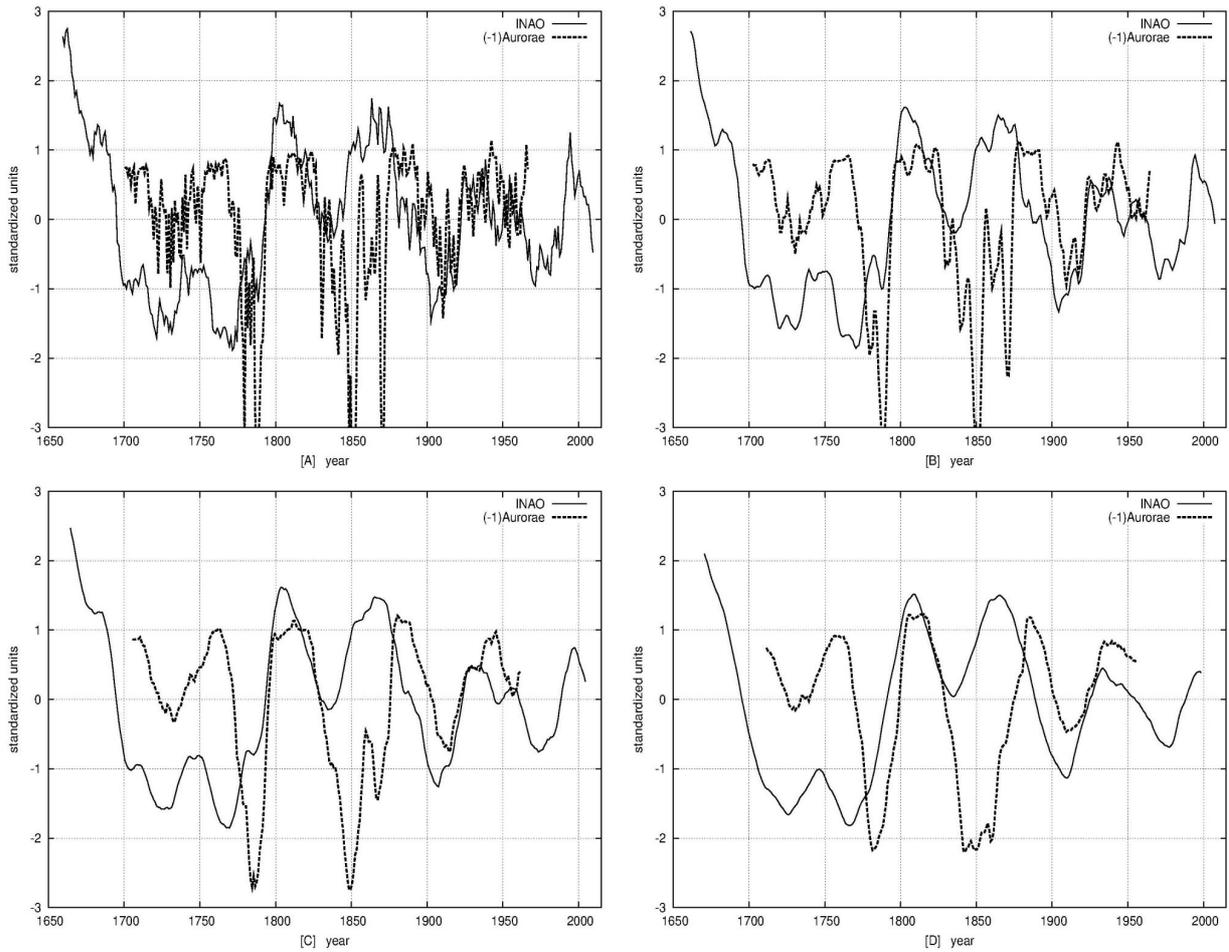

Figure 6. Time plot of yearly detrended values of INAO and Aurorae: (a) raw values, (b) smoothed according to a 5-yr running mean, (c) smoothed according to a 11-yr running mean, (d) smoothed according to a 23-yr running mean. The symbol (-1) before aurorae indicates that the record has been up-down flipped for a better visual comparison.

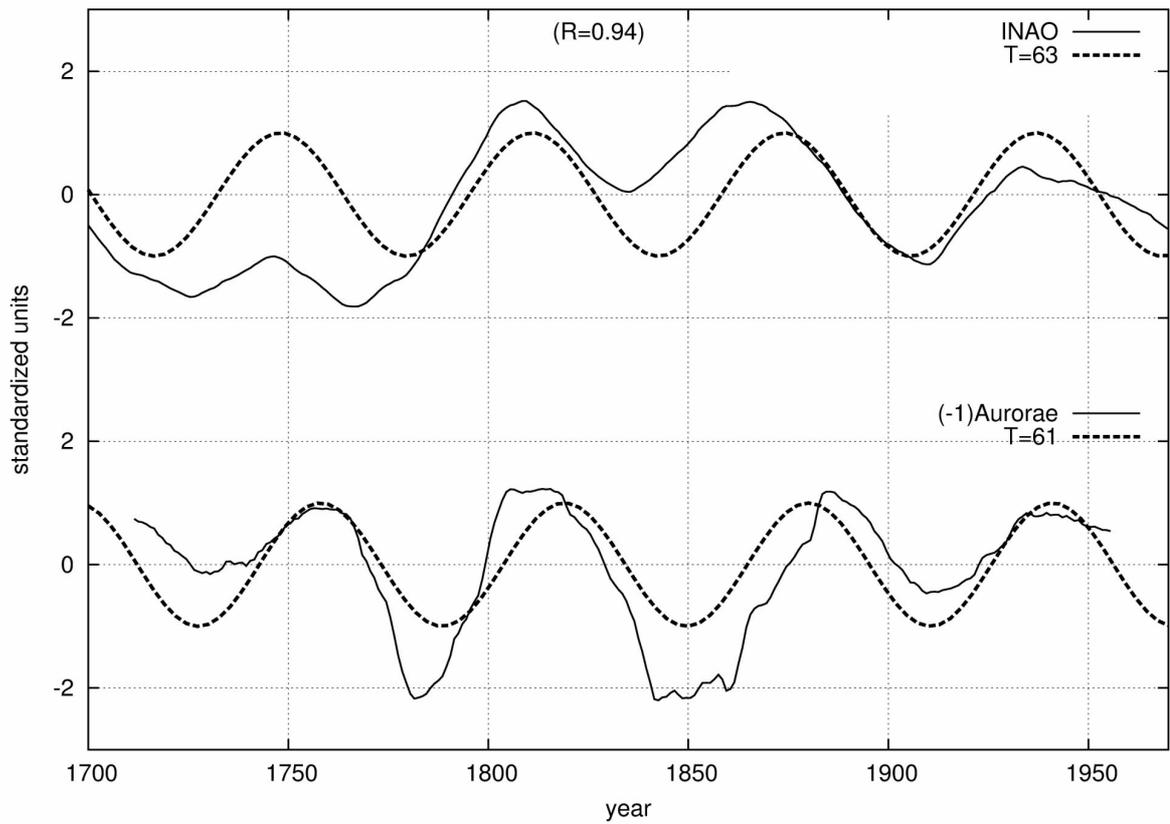

Figure 7. Time plot of yearly detrended values of INAO and Aurorae smoothed according to a 23-yr running mean. The two records are displaced for better visual. The records are fit with sinusoidal curves that show the existence of a ~60-year coherent oscillation in these records.

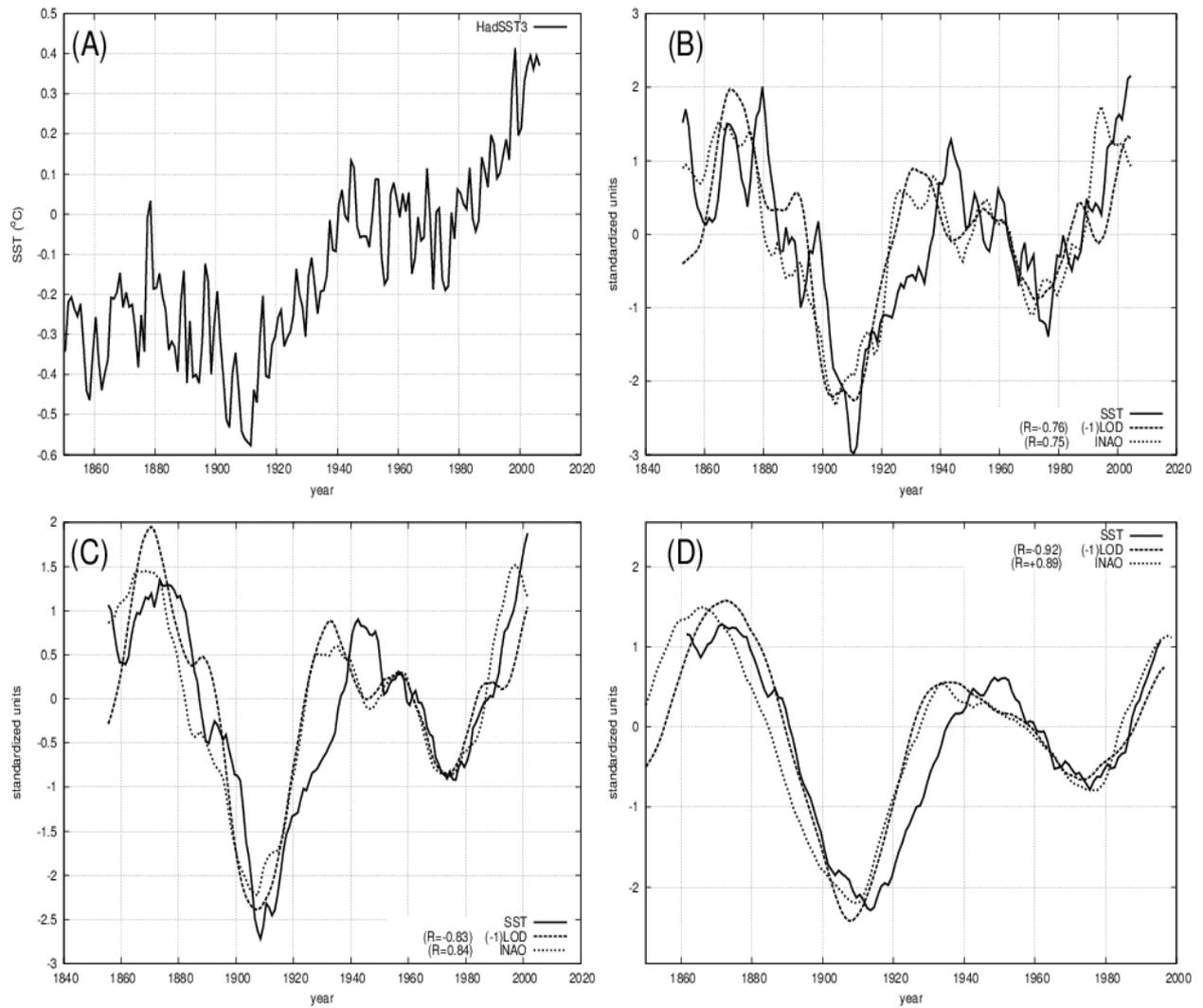

**Fig. 8** Time plot of yearly of HadSST3, LOD and INAO: **a** raw values of HadSST3, **b** smoothed according to a 5-year running mean, **c** smoothed according to an 11-year running mean, **d** smoothed according to a 23-year running mean.